\documentclass[aps,pre,twocolumn,superscriptaddress,showpacs,showkeys]{revtex4-1}
\usepackage{amsthm,amssymb,amsmath}
\usepackage{bm}
\usepackage[]{caption}
\usepackage{fancyhdr}
\usepackage{graphicx}
\usepackage{subfigure}
\usepackage{tikz}
\usepackage{hyperref}
\usepackage{ragged2e}

\usepackage{newtxtext,newtxmath}

\newcommand\ket[1]{{|#1\rangle}}

\newcommand\bbra[1]{{\langle\!\langle#1|}}
\newcommand\kket[1]{{|#1\rangle\!\rangle}}

\newcommand\opd{\mathbf{d}}
\newcommand\ope{\mathbf{e}}

\newcommand\dd{{\rm d}}
\newcommand\ee{{\rm e}}
\newcommand\ii{{\rm i}}
\newcommand\genfracz[2]{\genfrac{}{}{0pt}{}{#1}{#2}}

\newcommand\gammatwid{\widetilde\gamma}
\newcommand\deltatwid{\widetilde\delta}

\begin{document}

\title{Matrix product solution of a left-permeable two-species asymmetric exclusion process}

\author{Arvind Ayyer}
\email{arvind@iisc.ac.in}
\affiliation{Department of Mathematics, Indian Institute of Science, Bangalore - 560012, India.}

\author{Caley Finn}
\email{caley.finn@lapth.cnrs.fr }\thanks{Corresponding author.}
\affiliation{LAPTh,  CNRS - Universit{\'e} Savoie Mont Blanc, 9 chemin de Bellevue, BP 110, F-74941  Annecy-le-Vieux Cedex, France. }
\author{Dipankar Roy}
\email{dipankarroy@iisc.ac.in}
\affiliation{Department of Mathematics, Indian Institute of Science, Bangalore - 560012, India.}

\date{\today}

\begin{abstract}
We study a two-species partially asymmetric exclusion process where the left boundary is permeable for the
`slower' species but the right boundary is not. We find a matrix product solution for the stationary state,
and the exact stationary phase diagram for the densities and currents.  By calculating the density of each
species at the boundaries, we find further structure in the stationary phases.  In particular, we find that
the slower species can reach and accumulate at the far boundary, even in phases where the bulk density of
these particles approaches zero.
\end{abstract}



\maketitle

\section{Introduction}
Exclusion processes on finite lattices in contact with reservoirs are prototypical models of nonequilibrium
statistical mechanics. Although these models are defined by simple dynamical rules, they exhibit rich
phenomenology. Moreover, they have the property of being integrable, so that they can be analysed
rigorously~\cite{DerridaEHP93}. The simplest variant consists of a single type (or species) of particle and is
called the Asymmetric Simple Exclusion Process (ASEP). When the asymmetry is total (resp. partial), it is
called the TASEP (resp. PASEP). ASEPs with more than one kind of particle have found applications in recent
times in biological~\cite{simpson-et-al-2009,chou-mallick-2011} and chemical~\cite{bruna-chapman-2012}
systems.

While the most general variant of the single-species ASEP has an integrable structure, this is no longer true
even if there are two species of particles. In earlier work, progress has been made on understanding
two-species exclusion process with boundaries. Evans, Foster, Godr\'eche and Mukamel showed that a special
choice of boundary interactions exhibits spontaneous symmetry breaking~\cite{evans-et-al-1995}. Arita
considered a semipermeable TASEP, where the slower species (also known as second-class particles) were trapped
in the system, and determined the phase diagram~\cite{arita-2006}. Detailed properties of the phase diagram of
this model were analysed by Ayyer, Lebowitz and Speer~\cite{als-2009}.  The latter also studied some two-species models
whose phase diagram was determined using coloring techniques~\cite{AyyerLS12}.  Uchiyama, in a remarkable
paper, generalized the semipermeable TASEP to the semipermeable ASEP  by using considerably more
sophisticated techniques, and derived the phase diagram for the semipermeable PASEP \cite{Uchiyama08}.  More recently, integrable
two-species models were classified by Crampe, Mallick, Ragoucy and Vanicat~\cite{CrampeMRV15}.  The detailed
solution for one of the new integrable models discovered there was given by Crampe, Evans, Mallick, Ragoucy
and Vanicat~\cite{CrampeEMRV16}.  In a different direction, combinatorial and algebraic properties of
two-species exclusion processes have been studied by Duchi and Schaeffer~\cite{duchi-schaeffer-2005}, Corteel,
Mandelshtam and Williams~\cite{corteel-mandelshtam-williams-2015}, Mandelshtam and
Viennot~\cite{mandelshtam-viennot-2015} and Cantini~\cite{Cantini15}.

In this article, we focus on one of the integrable classes of two-species exclusion processes, where the slower particle
can only enter and exit from the left boundary. We call this the {\em left-permeable two-species ASEP}.
We begin with the preliminaries in Sec.~\ref{sec:prelim}.
We derive the matrix product solution for the stationary distribution in Sec.~\ref{sec:stationary} 
and find a representation of the matrix algebra in Sec.~\ref{sec:rep algebra}.
We find the phase diagram of the model in the thermodynamic limit
and derive formulas for the densities and current in all phases in Sec.~\ref{sec:phase}.
We end by computing the enriched phase diagrams for two different order parameters in Sec.~\ref{sec:21densities}:
the density of the slower particle at the last site, and the difference of bulk and boundary densities for both species.
We note that a large part of these calculations are generalizations of Uchiyama's techniques~\cite{Uchiyama08}.

\section{Preliminaries	\label{sec:prelim}}

\subsection{ Definition of the model}
The two-species ASEP describes particles hopping on a one dimensional lattice.  We consider
a finite lattice of length $L$ where each lattice site is either empty, or occupied by a single particle of
species 1 or 2.  Particles move along the lattice by exchanging places with their
immediate neighbours.  We can consider an empty site as a particle of species 0,
and then specify a lattice configuration by the tuple $\bm\tau = (\tau_1, \ldots, \tau_L)$, $\tau_i \in
\{ 0, 1, 2\}$.  In the bulk, exchanges between neighbouring particles occur with rates
\begin{equation}
\tau_i \tau_{i+1} \to \tau_{i+1} \tau_i \text{ with rate}
\begin{cases}
p, & \tau_i > \tau_{i+1},
\\
q, & \tau_i < \tau_{i+1}.
\end{cases}
\label{eq:bulkRates}
\end{equation}
We will take $p > q$ so that a particle of species $j$ moves preferentially to the right ahead of
all species $i < j$.

At the boundaries, we allow particles to enter and exit with the following rates:
\begin{itemize}
	\item Left boundary:
	\begin{equation}
	\begin{aligned}
	& 0 \to 1 \text{ with rate } \gamma,
	\\
	& 0, 1 \to 2 \text{ with rate } \alpha,
	\\
	& 2 \to 1 \text{ with rate } \gammatwid.
	\end{aligned}
	\label{eq:leftRates}
	\end{equation}
	
	\item Right boundary:
	\begin{equation}
	\begin{aligned}
	& 0 \to 2 \text{ with rate } \delta,
	\\
	& 2 \to 0 \text{ with rate } \beta.
	\end{aligned}
	\label{eq:rightRates}
	\end{equation}
\end{itemize}
The rate $\gammatwid$ is fixed as
\begin{equation}
\gammatwid = \frac{\alpha + \gamma + q - p}{\alpha + \gamma}\gamma.
\label{eq:gammatwid}
\end{equation}
The other rates $p, q, \alpha, \gamma, \beta, \delta$ can be arbitrary positive real numbers, subject to the constraint
\begin{equation}
\alpha + \gamma + q - p \ge 0,
\label{eq:mixingConstraint}
\end{equation}
so that $\gammatwid$ is not negative.  With this choice of rates, the model is integrable
\cite{CrampeMRV15,CrampeFRV16}.  Although we will not make direct use of the machinery of integrability, we
will see that the constraint in Eq.~\eqref{eq:gammatwid} also arises directly from the matrix product algebra
approach.

The boundary rates (Eq.~\eqref{eq:leftRates} and \eqref{eq:rightRates}) allow species 2 to enter and exit at both
boundaries.  With $q < p$ there will be a non-zero current of these particles from left to right, and so the
system is out of equilibrium.  In contrast, species 1 can only enter exit at the left boundary, and so
although this species is driven in the bulk, its net current will be zero.  Because species 1 is blocked by
the right boundary but not by the left, we say that this model is \emph{left-permeable}.

If instead of the left boundary rates (Eq.~\eqref{eq:leftRates}), we take
\begin{equation*}
\begin{aligned}
& 0 \to 2 \text{ with rate } \alpha,
\\
& 2 \to 0 \text{ with rate } \gamma,
\end{aligned}
\end{equation*}
(keeping the right-boundary rates in Eq.~\eqref{eq:rightRates}), species 1 is trapped on the lattice.  
We call this the semipermeable ASEP. Again, the net current of species 1 is zero, but in addition the \emph{number} of particles of species 1 is fixed. Thus the system decomposes into sectors according to the number of particles of species 1 on the lattice. The stationary state for this semipermeable model was found in matrix product form first for $q=\gamma=\delta=0$ \cite{arita-2006} and then in general \cite{Uchiyama08}. Later it was also studied through a Koornwinder polynomial approach \cite{Cantini15}. In this work we will follow the approach of \cite{Uchiyama08} and show that it can also be applied to the left-permeable model.

\subsection{Markov process formulation}
The models we have described are in fact continuous time Markov processes, which
can be specified formally by giving the transition matrix.  To do so, we must specify a basis.  To a site $i$,
with state given by $\tau_i$, we associate the standard basis vector $\ket{\tau_i} \in \mathbb{C}^3$, that is
\begin{equation*}
 \ket{0} = \begin{pmatrix} 1 \\ 0 \\ 0 \end{pmatrix},
 \qquad
 \ket{1} = \begin{pmatrix} 0 \\ 1 \\ 0 \end{pmatrix},
 \qquad
\ket{2} = \begin{pmatrix} 0 \\ 0 \\ 1 \end{pmatrix}.
\end{equation*}
Then the lattice configuration is given by a vector $\ket{\bm\tau}
\in (\mathbb{C}^3)^{\otimes L}$,
\begin{equation*}
\ket{\bm\tau} = \ket{\tau_1, \ldots, \tau_L} = \ket{\tau_1} \otimes \ldots \otimes \ket{\tau_L}.
\end{equation*}
The rates at which neighbouring particles exchange places (Eq.~\eqref{eq:bulkRates})
are encoded in the local transition  matrix $w \in \mathbb{C}^3 \otimes \mathbb{C}^3$,
\begin{equation}
w = \begin{pmatrix}
0 & 0 & 0 & 0 & 0 & 0 & 0 & 0 & 0
\\
0 & -q & 0 & p & 0 & 0 & 0 & 0 & 0
\\
0 & 0 & -q & 0 & 0 & 0 & p & 0 & 0
\\
0 & q & 0 & -p & 0 & 0 & 0 & 0 & 0
\\
0 & 0 & 0 & 0 & 0 & 0 & 0 & 0 & 0
\\
0 & 0 & 0 & 0 & 0 & -q & 0 & p & 0
\\
0 & 0 & q & 0 & 0 & 0 & -p & 0 & 0
\\
0 & 0 & 0 & 0 & 0 & q & 0 & -p & 0
\\
0 & 0 & 0 & 0 & 0 & 0 & 0 & 0 & 0
\end{pmatrix},
\label{eq:wbulk}
\end{equation}
acting on the ordered basis,
\begin{equation*}
\{\ket{0,0}, \ket{0,1}, \ket{0,2},  \ket{1,0}, \ket{1,1}, \ket{1,2},\ket{2,0}, \ket{2,1}, \ket{2,2} \}. \nonumber 
\end{equation*}
The boundary rates (Eq.~\eqref{eq:leftRates} and \eqref{eq:rightRates}) are encoded (respectively) by matrices
$B$, $\overline{B} \in \mathbb{C}^3$:
\begin{equation}
B = \begin{pmatrix}
-\alpha - \gamma & 0 & 0
\\
\gamma & -\alpha & \gammatwid
\\
\alpha & \alpha & -\gammatwid
\end{pmatrix},
\ 
\overline{B} = \begin{pmatrix}
-\delta & 0 & \beta
\\
0 & 0 & 0
\\
\delta & 0 & -\beta
\end{pmatrix}.
\label{eq:mixingBoundaries}
\end{equation}
The complete transition matrix is then given by the sum of local matrices
\begin{equation}
M = B_1 + \sum_{i=1}^{L-1} w_{i,i+1} + \overline{B}_L.
\label{eq:Mcomplete}
\end{equation}
The subscripts indicate the sites on which each matrix acts.  That is,
\begin{eqnarray}
B_1 &=& B \otimes I^{(L-1)},
\nonumber\\ 
w_{i,i+1} &=& I^{(i-1)} \otimes w \otimes I^{(L-i-1)},
\nonumber\\ 
\overline{B}_L &=& I^{(L-1)} \otimes \overline{B}, \nonumber
\end{eqnarray}
where $I^{(k)}$ is the identity matrix on the $k$-fold tensor product of $\mathbb{C}^3$.

Writing $P_{\bm\tau}(t)$ for the probability of a configuration $\bm\tau$ at time $t$, the time evolution is
determined by the master equation
\begin{eqnarray}
\frac{\dd}{\dd t} \ket{P(t)} &=& M \ket{P(t)}, \nonumber \\
 \text{where} \qquad  
\ket{P(t)} &=& \sum_{\bm\tau} P_{\bm\tau}(t) \ket{\bm\tau}. \nonumber
\end{eqnarray}
At late times, the system converges to the stationary distribution of the process given by the normalized
eigenvector of $M$ with eigenvalue $0$.  That is, with
\begin{equation*}
\ket{\Psi} = \sum_{\bm\tau} \psi_{\bm\tau} \ket{\bm\tau},
\qquad
M \ket{\Psi} = 0,
\end{equation*}
the stationary distribution is
\begin{equation*}
\ket{P_{\text{stat}}} = \frac{1}{Z_L} \ket{\Psi},
\qquad
Z_L =\sum_{\bm\tau} \psi_{\bm\tau}.
\end{equation*}
We will see later that the normalisation $Z_L$ plays a role
analogous to that of the partition function in equilibrium statistical mechanics. 
We will, with some abuse of terminology, refer to $Z_L$ as the partition function throughout the paper.

From the stationary distribution we can compute the density of species $k = 1, 2$ at site $i$
\begin{equation*}
\rho_i^{(k)} = \frac{1}{Z_L} \sum_{\substack{\bm\tau \\ \tau_i = k}} \psi_{\bm\tau},
\end{equation*}
and we write $\rho^{(k)}$ for the density averaged across the lattice.  We can also compute the current
$J^{(2)}$, of species 2: the probability per unit time that a particle of species 2 crosses a fixed point on the
lattice (see Eq.~\eqref{eq:J2}).  Recall that the net current of species 1 is zero.

There is also a \emph{right-permeable} two-species model analogous to the left-permeable model, with boundary
matrices
\begin{equation*}
B = \begin{pmatrix}
-\alpha & 0 & \gamma
\\
0 & 0 & 0
\\
\alpha & 0 & -\gamma
\end{pmatrix},
\qquad
\overline{B} = \begin{pmatrix}
-\deltatwid & \beta & \beta
\\
\deltatwid & -\beta & \delta
\\
0 & 0 & -\beta - \delta
\end{pmatrix},
\end{equation*}
with
\begin{equation*}
\deltatwid = \frac{\beta + \delta + p - q}{\beta + \delta}\delta,
\end{equation*}
and the same bulk matrix (Eq.~\eqref{eq:wbulk}).  If we write the unnormalized stationary state vector for the
left-permeable model as
	\begin{equation*}
	\ket{\Psi^{\text{left}}(\alpha, \beta, \gamma, \delta; p, q)}
	=
	\sum_{\bm\tau} \psi^{\text{left}}_{\bm\tau}(\alpha, \beta, \gamma, \delta; p, q)\ket{\bm\tau},
	\end{equation*}
the weights for the right-permeable model are given by
	\begin{equation*}
		\psi^{\text{right}}_{\bm\tau}(\alpha, \beta, \gamma, \delta; p, q)
		=
		\psi^{\text{left}}_{\bm{\widetilde\tau}}(\beta, \alpha, \delta, \gamma; q, p),
		\quad
		\widetilde\tau_i = 2 - \tau_{L-i+1}.
	\end{equation*}
Note for the right-permeable model, we take $q > p$.  Taking $q < p$ would correspond to a reverse-biased
regime, where the boundary rates oppose the preferred direction of flow in the bulk \cite{BlytheECE00,deGierFS11}.
\subsection{Stationary phase diagram of the semipermeable ASEP}
\label{sec:ASEPphases}

We first review the key features of the stationary state of the semipermeable ASEP, since this will be useful for us later.
The full phase diagram was computed in \cite{Uchiyama08} using a matrix product algebra
\cite{DerridaEHP93,BlytheE07}.  The phase diagram has the same general structure as that of the single species
ASEP \cite{Sandow94,UchiyamaSW04}.

The key quanties of interest are the current and average
density of the particles of species 2, $J^{(2)}$ and $\rho^{(2)}$, respectively.  As species 1 is trapped on
the lattice, the average density $\rho^{(1)}$ is a fixed parameter.  The other parameters determining the
phases of the system are expressed as the combinations of rates
\begin{equation}
\label{eq:abcd general}
a = \kappa^+_{\alpha,\gamma},
\ \
c = \kappa^-_{\alpha,\gamma},
\ \
b = \kappa^+_{\beta,\delta},
\ \
d = \kappa^-_{\beta,\delta},
\end{equation}
where
	\begin{equation*}
		\kappa_{u, v}^\pm = \frac{1}{2 u}\left(
		p - q - u + v \pm \sqrt{(p - q - u + v)^2 + 4 u v}
		\right).
\end{equation*}
This parameterisation satisfies $a, b \ge 0$, and with $p > q$, $-1 < c, d \le 0$. 
\begin{figure}[ht]
	\centering
	\includegraphics[width=0.4\textwidth]{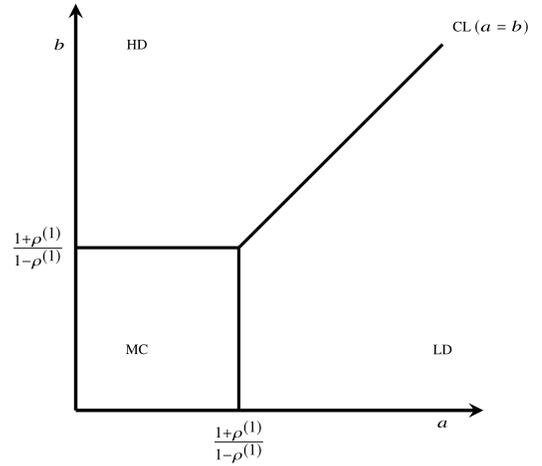}
	\caption{
		Stationary phase diagram of the semipermeable two-species ASEP.  The density of species 1, $\rho^{(1)}$
		is a fixed parameter in this system.
	}
	\label{fig:trappedphases}
\end{figure}
The phase diagram, shown in Fig.~\ref{fig:trappedphases}, depends only on $a$, $b$, and the density
$\rho^{(1)}$.  We name the phases according to the behavior of species 2.  The current and average
density in these phases are:
\begin{itemize}
	\item
	Maximum current (MC) phase: $a, b < (1 + \rho^{(1)})/(1 - \rho^{(1)})$, with
	\begin{equation*}
	\rho^{(2)} = \frac{1 - \rho^{(1)}}{2}, \qquad J^{(2)} = \frac{(p-q)(1 - {\rho^{(1)}}^2)}{4};
	\end{equation*}
	
	\item
	Low density (LD) phase: $a > (1 + \rho^{(1)})/(1 - \rho^{(1)})$, $a > b$, with
	\begin{equation*}
	\rho^{(2)} = \frac{1}{1 + a}, \qquad J^{(2)} = \frac{(p - q) a}{(1 + a)^2};
	\end{equation*}
	
	\item
	High density (HD) phase: $b > (1 + \rho^{(1)})/(1 - \rho^{(1)})$, $b > a$, with
	\begin{equation*}
	\rho^{(2)} = \frac{b}{1 + b} - \rho^{(1)}, \qquad J^{(2)} = \frac{(p - q) b}{(1 + b)^2}.
	\end{equation*}
\end{itemize}

The line $a = b > 1$ separating the high and low density phases is called the coexistence line (CL).  Here
both high and low density domains can exist on the lattice.  This situation also exists for the single species
ASEP, and is described by a domain wall model \cite{KolomeiskySKS98}. 

\section{Stationary state for the left-permeable two-species ASEP}
\label{sec:stationary}

In this section we give a matrix product algebra and representation for the left-permeable two-species ASEP with
boundary matrices (Eq.~\eqref{eq:mixingBoundaries}).  From this point on, we fix the rightwards hopping rate to $p =
1$.  We can do this without loss of generality, as it corresponds to rescaling the unit of time.

\subsection{Matrix product algebra}
To write the stationary probabilities in matrix product form, define two vectors,
\begin{equation*}
X = \begin{pmatrix}
E \\ A \\ D
\end{pmatrix},
\qquad
x = \begin{pmatrix}
-1 \\ 0 \\ 1
\end{pmatrix}.
\end{equation*}
The entries of $X$ ($E$, $A$, $D$) are matrices in some auxilliary space.  We take
$\bbra{W}$, $\kket{V}$ as left and right vectors in this space that contract the matrices to give scalar
values.  We write the unnormalized stationary vector as
\begin{equation}
	\ket{\Psi} = \bbra{W} X \otimes \ldots \otimes X \kket{V},
	\label{eq:mprodPsi}
\end{equation}
so that
\begin{equation}
	\psi_{\bm\tau} = \bbra{W} X_{\tau_1} \ldots X_{\tau_L} \kket{V}.
	\label{eq:mprodcomponent}
\end{equation}
That is to say, in every configuration, the occurence of $0$ is represented by $E$, $1$ by $A$, and $2$ by $D$.
In order for this construction to give the stationary state, it is sufficient to find matrices $E, A, D$ and vectors $\bbra W, \kket V$ for
which the following conditions hold:
\begin{equation}
	\begin{aligned}
		B \bbra{W} X &= \bbra{W} x,
		\\
		w X \otimes X &= -x \otimes X + X \otimes x,
		\\
		\overline{B} X \kket{V} &= -x \kket{V}.
		\label{eq:mprodrels}
	\end{aligned}
\end{equation}
If we apply the transition matrix $M$ of form Eq.~\eqref{eq:Mcomplete} to $\ket{\Psi}$ given by
Eq.~\eqref{eq:mprodPsi}, these relations cause the bulk sum to telescope to two terms, which are cancelled by the
left and right boundary terms (see \cite{DerridaEHP93,BlytheE07,CrampeRV14} where this is discussed in more
detail).  It is important to note that we must also show that relations (Eq.~\eqref{eq:mprodrels}) are consistent.
We will do this, in the usual manner, by giving explicit matrices $E$, $D$, $A$, and boundary vectors $\bbra{W}$,
$\kket{V}$ that satisfy the relations in Eq.~\eqref{eq:mprodrels}.  In fact, we will see that all we require is a
representation of same algebra as used for the semipermeable model in \cite{Uchiyama08}, and we review that
representation in Sec.~\ref{sec:rep algebra}.

With $w$ given by Eq.~\eqref{eq:wbulk}, the bulk relations implied by Eq.~\eqref{eq:mprodrels} are
\begin{equation}
\begin{aligned}
& D E - q E D = D + E,
\\
& 
A E - q E A = A,
\\
& D A - q A D = A.
\end{aligned}
\label{eq:bulkrels}
\end{equation}
And with boundary matrices (Eq.~\eqref{eq:mixingBoundaries}), the boundary relations are
\begin{equation}
\begin{aligned}
& (\alpha + \gamma) \bbra{W} E = \bbra{W},
\\
& \gamma \bbra{W} E  - \alpha \bbra{W} A + \gammatwid \bbra{W} D = 0,
\\
& -\delta E \kket{V} + \beta D \kket{V} = \kket{V}.
\end{aligned}
\label{eq:boundaryrels}
\end{equation}
Using these relations, any expression of the form of Eq.~\eqref{eq:mprodcomponent} can be reduced to a scalar multiple of
$\langle\!\langle W | V \rangle\!\rangle$.  For small system sizes, we can compute the stationary weights in this way.
Checking that the computed vector is in fact the eigenvector of the transition matrix in Eq.~\eqref{eq:Mcomplete} with
eigenvalue zero, we find that  it is \emph{necessary} that $\gammatwid$
takes on its integrable value (Eq.~\eqref{eq:gammatwid}).  We will show that this constraint on the parameters is
also sufficient by giving an explicit representation of this algebra.

The bulk relations (Eq.~\eqref{eq:bulkrels}) are the same as those from \cite{Uchiyama08} for the semipermeable ASEP.  
Following \cite{Uchiyama08}, we express the bulk relations in
terms of matrices $\ope$, $\opd$, satisfying the $q$-deformed oscillator algebra \cite{Sandow94},
\begin{equation}
\opd \ope - q \ope \opd = 1 - q.
\label{eq:qoscalg}
\end{equation}
Then
\begin{equation*}
\begin{aligned}
& D = \frac{1}{1-q}(1 + \opd),
\qquad
E = \frac{1}{1-q}(1 + \ope),
\\
& A = \lambda (D E - E D) = \frac{\lambda}{1-q} (1 - \ope \opd),
\end{aligned}
\end{equation*}
satisfy the bulk algebra, with $\lambda$ a free parameter. We will write the boundary relations as
\begin{equation}
\begin{aligned}
& \bbra{W} \ope  + a c \bbra{W} \opd = (a + c) \bbra{W},
\\
& \opd \kket{V} + b d \ope \kket{V} = (b+d)\kket{V}.
\end{aligned}
\label{eq:boundaryalg}
\end{equation}
This is the form used for the semipermeable ASEP \cite{Uchiyama08}, and also the single species ASEP
\cite{UchiyamaSW04}.  In both these cases the parameters $a$, $b$, $c$, $d$ are those appearing in the
stationary state of the ASEP (see Eq.~\eqref{eq:abcd general}).

The \emph{three} boundary relations (Eq.~\eqref{eq:boundaryrels}) for the left-permeable
two-species ASEP reduce to \emph{two} relations of the form in 
Eq.~\eqref{eq:boundaryalg} if we fix $\lambda = \gamma/\alpha$ and $\gammatwid$ at the value given in Eq.~\eqref{eq:gammatwid}.  
The parameters $a$, $b$, $c$, $d$ are given by
\begin{equation}
a = 0, \;
c = \frac{1 - q - \alpha - \gamma}{\alpha + \gamma}, \;
b = \kappa_{\beta,\delta}^+, \;
d = \kappa_{\beta,\delta}^-.
\label{eq:abcd}
\end{equation}
With the constraint
Eq.~\eqref{eq:mixingConstraint}, we can write
\begin{equation}
	a = \kappa_{\alpha+\gamma,0}^+, \ c = \kappa_{\alpha+\gamma,0}^-, \ \mathrm{and} \
	-1 < c \le 0,
	\label{eq:cconstraint}
\end{equation}
where the lower bound assumes that $q < 1$.

Representations of the algebra (Eq.~\eqref{eq:qoscalg} and \eqref{eq:boundaryalg}) are well known, and
in Sec.~\ref{sec:representation} we will give the explicit form of the representation used in
\cite{Uchiyama08}. Since we know that a representation exists, the matrix product relations for the left-permeable two-species ASEP are consistent, and can be used to calculate the stationary state. But first, we describe the main physical quantities of interest, and how they are calculated.

\subsection{Physical quantities}
The stationary probabilities are obtained by normalising the stationary weights (Eq.~\eqref{eq:mprodcomponent}).  Thus
\begin{eqnarray}
P_{\bm\tau} &=& \frac{1}{Z_L}\bbra{W}X_{\tau_1} \ldots X_{\tau_L}\kket{V},
\label{eq:Pstat} \\
\mathrm{with} \ Z_L &=& \bbra{W} C^L \kket{V}, \quad C = E + D + A. \nonumber
\end{eqnarray}
The current of type 2 particles is given by
	\begin{align}
		J^{(2)} &= \frac{1}{Z_L} \bbra{W} C^{i-1}\left( D E - q E D + D A - q A D\right) C^{L-i-1} \kket{V} \nonumber \\
		&= \frac{Z_{L-1}}{Z_L},
		\label{eq:J2} \end{align}
which is independent of position, $i$.  The net current of type 1 particles must be zero as they can only enter
at the left boundary.  Indeed, computing with the matrix product algebra, we find
	\begin{equation*}
	\begin{aligned}
		J^{(1)} =& \frac{1}{Z_L} \bbra{W} C^{i-1}\left( A E - q E A + q A D - D A\right) C^{L-i-1} \kket{V} \\
		=& 0.
	\end{aligned}
	\end{equation*}
We would also like to compute the average density of species $k = 1,2$, which is given by
\begin{equation}
\rho^{(k)} = \frac{1}{L}\frac{1}{Z_L}\sum_{i=1}^L \bbra{W}C^{i-1} X_k C^{L-i}\kket{V}.
\label{eq:rhok}
\end{equation}
To achieve this, we define
\begin{equation}
Z_L(\xi^2, \zeta) = \bbra{W} \left(E + \xi^2 D + \zeta A\right)^L \kket{V},
\label{eq:Zpart}
\end{equation}
which plays the role of a partition function with fugacities $\xi^2, \zeta$ for type 1 and 2 particles respectively.  
Then
\begin{equation}
\begin{aligned}
\rho^{(1)} & = \frac{1}{L} \frac{\partial}{\partial \zeta} \log Z_L(\xi^2, \zeta)\big|_{\xi^2=\zeta=1},
\\
\rho^{(2)} & = \frac{1}{L} \frac{\partial}{\partial \xi^2} \log Z_L(\xi^2, \zeta)\big|_{\xi^2=\zeta=1}.
\end{aligned}
\label{eq:rhokdiff}
\end{equation}
In order to compute the partition function defined in Eq.~\eqref{eq:Zpart}, it will be convenient to rewrite it as
\begin{equation}
	\begin{aligned}
		Z_L(\xi^2, \zeta)
		=&
		\left(\frac{\xi}{1-q}\right)^L\\
		&\times\bbra{W}\left(
		\xi^{-1} + \xi + \bar{\ope} + \bar{\opd} + (1 - q) \bar{\zeta} A
		\right)^L\kket{V},
		\label{eq:Zpartbar}
	\end{aligned}
\end{equation}
with
\begin{equation*}
\bar{\ope} = \xi^{-1} \ope,
\qquad
\bar{\opd} = \xi \opd,
\qquad
\bar{\zeta} = \zeta \xi^{-1}.
\end{equation*}

The rescaled generators $\bar{\ope}$, $\bar{\opd}$ satisfy the same $q$-oscillator algebra (Eq.~\eqref{eq:qoscalg}).
Defining also
\begin{equation*}
\bar{a} = \xi^{-1} a,
\quad
\bar{c} = \xi^{-1} c,
\quad
\bar{b} = \xi b,
\quad
\bar{d} = \xi d,
\end{equation*}
the boundary relations for the rescaled generators are obtained by putting bars over the boundary parameters
$a$, $b$, $c$, $d$ in Eq.~\eqref{eq:boundaryalg}.  Thus, given a representation of the original algebra, we obtain
a representation of the scaled algebra, simply by replacing the boundary parameters by their barred versions.

\section{Representation of the algebra and the partition function}
\label{sec:rep algebra}

For the representation of the algebra we use exactly that from \cite{Uchiyama08}, but with parameters
specialized differently.  We then review how this is used to find an integral form for the partition function.

\subsection{Continuous big $q$-Hermite polynomials}
To give the representation of the algebra, we must first introduce certain notation from the `$q$-calculus' \cite{GasperR90,Koekoek10}.  The $q$-shifted factorial is given by
\begin{equation*}
(a_1, \ldots, a_s; q)_n = \prod_{r=1}^s (a_r; q)_n,
\end{equation*}
where
\begin{align*}
(a; q)_n &= \prod_{k=0}^{n-1}(1 - a q^k)= (1-a) (1-aq) \cdots (1-aq^{n-1}),
\end{align*}
valid also for $n \to \infty$ when $q < 1$. The basic hypergeometric series is given by
\begin{equation*}
	\begin{aligned}
		_r \phi_s\left[\genfracz{a_1, \ldots, a_r}{b_1, \ldots, b_s} \Bigg| q, z \right]
		=
		\sum_{k=0}^\infty &\frac{(a_1, \ldots, a_r; q)_k}{(q, b_1, \ldots, b_s; q)_k} \\
		& \times\left(
		(-1)^k q^{\binom{k}{2}}
		\right)^{1+s-r} z^k.
	\end{aligned}
\end{equation*}
Following \cite{Uchiyama08}, we define
\begin{equation*}
F_n(u, v; \lambda) = \sum_{k=0}^n \frac{(q;q)_n}{(q;q)_k (q;q)_{n-k}}(\lambda u;q)_k v^k u^{n-k},
\end{equation*}
which satisfies the recurrence relation
\begin{equation*} \begin{aligned}
		F_{n+1}(u,v;\lambda)
		+ \lambda u v q^n F_n(u,v;\lambda) + &(1 - q^n) u v F_{n-1}(u,v;\lambda) \\
		&  = (u + v)F_n(u,v;\lambda), \end{aligned}
\end{equation*}
with $F_{-1} = 0$ and $F_0 = 1$.  Specialisation of the parameters $u,v$ gives the {\em continuous big $q$-Hermite polynomial}
\cite{Koekoek10},
\begin{equation*}
H_n(\cos\theta;\lambda|q) = F_n(\ee^{\ii \theta}, \ee^{-\ii \theta}; \lambda).
\end{equation*}
For $\lambda$ real and $|\lambda| < 1$, $H_n(\cos\theta; \lambda|q)$ satisfies the orthogonality relation
	\begin{equation} \begin{aligned}
		\int_0^\pi \frac{\dd \theta}{2\pi}
		w(\cos\theta; \lambda) \, H_m(\cos\theta; \lambda|q) \, & H_n(\cos\theta; \lambda|q) \\
		& = (q; q)_n \delta_{mn},
		\label{eq:BigHorthog} \end{aligned}
	\end{equation}
with
\begin{equation}
w(\cos\theta; \lambda)
=
w(\ee^{\ii\theta}, \ee^{-\ii\theta}; \lambda)
=
\frac{(q, \ee^{2\ii\theta}, \ee^{-2\ii\theta}; q)_\infty}
{(\lambda \ee^{\ii\theta}, \lambda \ee^{-\ii\theta}; q)_\infty}.
\end{equation}
Alternatively, we can write Eq.~\eqref{eq:BigHorthog} as the contour integral
	\begin{align}
		\oint \frac{\dd z}{4\pi \ii z}
		w(z, z^{-1}; \lambda) \, H_m\left(\frac{z + z^{-1}}{2}; \lambda|q\right) &
		 H_n\left(\frac{z + z^{-1}}{2}; \lambda|q\right) \nonumber \\
		& = (q; q)_n \delta_{mn},
		\label{eq:BigHorthogContour} 
		\end{align}
where the contour of integration is the unit circle.  The orthogonality condition for $\lambda > 1$ is
obtained from Eq.~\eqref{eq:BigHorthogContour} by deforming the contour of integration: such that the origin and
all poles at $\lambda q^k$ are included, and all poles at $\lambda^{-1} q^{-k}$ are excluded, with
$k = 0, 1, 2, \ldots$.

We will need the $q$-Mehler-type sum formula given in \cite{Uchiyama08} (see also \cite{Bressoud80})
for $|\tau u|, |\tau v | < 1$:

\begin{equation}
\begin{aligned}
\sum_{n=0}^\infty \frac{\tau^n}{(q; q)n} H_n(\cos\theta; \lambda | q) \, F_n(u, v; 0)
= \Theta(\cos \theta; \tau u, \tau v | \lambda),
\label{eq:sumHF}
\end{aligned}
\end{equation}
where
\begin{equation}
\begin{aligned}
\Theta(\cos \theta; u, v | \lambda)
& \equiv
\Theta(\ee^{\ii\theta}, \ee^{-\ii\theta}; u, v | \lambda)
\\ 
&
 =
\frac{(\lambda u, \lambda v; q)_\infty}
{(u \ee^{\ii\theta}, u \ee^{-\ii\theta},
	v \ee^{\ii\theta}, v \ee^{-\ii\theta}; q)_\infty} \\
& \quad \times 
{_2}\phi_2\left[\genfracz{\lambda \ee^{\ii \theta}, \lambda \ee^{-\ii \theta}}
{\lambda u, \lambda v} \Bigg| q, u v \right].
\end{aligned}
\end{equation}
For the model we consider, we will need to take $u = 0$ (or equivalently $v = 0$), and can do this by taking
the limit $u \to 0$.  For convenience, we will write
\begin{equation*}
\begin{aligned}
& F_n(0, v; \lambda) = \lim_{u \to 0} F_n(u, v; \lambda) = v^n, \quad \text{and}
\\
& {_2}\phi_2\left[\genfracz{\lambda \ee^{\ii \theta}, \lambda \ee^{-\ii \theta}}
{0, \lambda v} \Bigg| q, 0 \right]
=
\lim_{u \to 0} {_2}\phi_2\left[\genfracz{\lambda \ee^{\ii \theta}, \lambda \ee^{-\ii \theta}}
{\lambda u, \lambda v} \Bigg| q, u v \right]
= 1.
\end{aligned}
\end{equation*}
Note also that if $|\tau u| > 1$ or $|\tau v | > 1$, the sum in Eq.~\eqref{eq:sumHF} is divergent as, for example, if
$|u| > |v|$, $F_n(u, v; 0) \sim u^n$ for large $n$.

\subsection{Representation}
\label{sec:representation}

The $q$-oscillator algebra (Eq.~\eqref{eq:qoscalg}) has a Fock space representation
\begin{equation}
	\begin{aligned}
		\opd &= \sum_{n=1}^\infty \sqrt{1 - q^n} \kket{n-1}\bbra{n}, \\
		\qquad
		\ope &= \sum_{n=0}^\infty \sqrt{1 - q^{n+1}} \kket{n+1}\bbra{n},
		\label{eq:edrep}
	\end{aligned}
\end{equation}
and therefore
\begin{equation*}
A = \frac{\lambda}{1-q}\sum_{n=0}^\infty q^n \kket{n}\bbra{n}.
\end{equation*}
Writing the boundary vectors as
\begin{equation}
\bbra{W} = \sum_{n=0}^\infty w_n \bbra{n},
\qquad
\kket{V} = \sum_{n=0}^\infty v_n \kket{n},
\label{eq:wvrep}
\end{equation}
then from the boundary relations (Eq.~\eqref{eq:boundaryalg}), the coefficients must satisfy
\begin{equation*}
\begin{aligned}
 \sqrt{(q; q)_{n+1}} w_{n+1} & - (a + c)\sqrt{(q; q)_n} w_n \\ 
	 + & a c (1 - q^n)  \sqrt{(q; q)_{n-1}} w_{n-1} = 0,
\\
 \sqrt{(q; q)_{n+1}} v_{n+1} & - (b+d)\sqrt{(q; q)_n} v_n \\ + & b d (1 - q^n)\sqrt{(q; q)_{n-1}} v_{n-1} = 0.
\end{aligned}
\end{equation*}
These recurrences are solved by taking
\begin{equation}
w_n = \frac{F_n(a, c; 0)}{\sqrt{(q; q)_n}}, \qquad v_n = \frac{F_n(b, d; 0)}{\sqrt{(q; q)_n}}.
\label{eq:wnvn}
\end{equation}
Note that as we have $a = 0$, $w_n$ has the simpler form
\begin{equation*}
w_n =  \frac{c^n}{\sqrt{(q; q)_n}}.
\end{equation*}
With Eqs.~\eqref{eq:edrep}, \eqref{eq:wvrep} and \eqref{eq:wnvn}, we have a representation of the algebra in
Eqs.~\eqref{eq:qoscalg} and \eqref{eq:boundaryalg}.

In order to compute the partition function $Z(\xi^2, \zeta)$, we start from the form Eq.~\eqref{eq:Zpartbar}.  The
trick is to find a solution of the eigenvalue equation
\begin{align}
(\bar{\ope} + \bar{\opd} + (1 - q) \bar{\zeta} A) & \kket{h(\cos\theta)} 
 = \  2 \cos\theta \kket{h(\cos\theta)}.
\label{eq:bulkevec}
\end{align}
Using the representation in Eq.~\eqref{eq:edrep} (which also gives a representation of the barred algebra), we find
\begin{equation*}
\kket{h(\cos\theta)}
= \sum_{n=0}^\infty \frac{H_n(\cos\theta; \lambda \zeta \xi^{-1}| q)}{\sqrt{(q; q)_n}} \kket{n}.
\end{equation*}
As $(\bar{\ope} + \bar{\opd} + (1 - q) \bar{\zeta} A)$ is a symmetric matrix, it has the transpose
$\bbra{h\cos\theta}$ as a right eigenvector with the same eigenvalue.  From the orthogonality condition
(Eq.~\eqref{eq:BigHorthog}), we then obtain
\begin{equation}
1 = \int_0^\pi \frac{\dd \theta}{2 \pi}
w(\cos\theta; \lambda \zeta \xi^{-1}) \, \kket{h(\cos\theta)}\bbra{h(\cos\theta)},
\label{eq:hhidentity}
\end{equation}
for $|\lambda \zeta \xi^{-1}| < 1$.  For the case $|\lambda \zeta \xi^{-1}| > 1$, we use the contour integral
form of the orthogonality condition, with the contour deformed as described below (Eq.~\eqref{eq:BigHorthogContour}).

\subsection{Partition function}
Following \cite{Uchiyama08,UchiyamaSW04}, we write the partition function in integral form.  Starting from the
partition function as given in Eq.~\eqref{eq:Zpartbar}, use Eq.~\eqref{eq:hhidentity} to insert the identity, then the
eigenvalue equation Eq.~\eqref{eq:bulkevec}, then finally the sum formula Eq.~\eqref{eq:sumHF}.  This gives the
integral form 
\begin{widetext}
	\begin{equation}
		Z_L(\xi^2, \zeta) = \int_0^\pi \frac{\dd \theta}{2 \pi}
		w(\cos\theta; \lambda \zeta \xi^{-1})
		\, 
		\Theta(\cos\theta; 0, \xi^{-1} c|\lambda \zeta \xi^{-1})
		\,
		\Theta(\cos\theta; \xi b, \xi d|\lambda \zeta \xi^{-1}) 
		\left( \frac{1 + \xi^2 + 2 \xi \cos\theta}{1 - q}\right)^L.
		\label{eq:ZpartTheta}
	\end{equation}
\end{widetext}
We have used the boundary vectors of the `barred' algebra to obtain this expression.  The form
Eq.~\eqref{eq:ZpartTheta} is
valid for $|\zeta \xi^{-1} \lambda|, |\xi b|, |\xi^{-1}c|, |\xi d| < 1$.  Recall also that for the
left-permeable model we have $a = 0$.

In fact, in our model $|c|, |d| < 1$, and we can take $\zeta$, $\xi$ arbitrarily close to $1$.
Thus we need only be concerned with the cases where $\lambda > 1$ or $b > 1$.  For these cases, we write the
partition function in Eq.~\eqref{eq:ZpartTheta} by changing to the variable $z = e^{\ii \theta}$ as 
\begin{widetext}
	\begin{equation}
		Z_L(\xi^2, \zeta) = \oint \frac{\dd z}{4 \pi \ii z}
		w(z, z^{-1}; \lambda \zeta \xi^{-1})
		\,
		\Theta(z, z^{-1}; 0, \xi^{-1} c | \lambda \zeta \xi^{-1})
		\,
		\Theta(z, z^{-1}; \xi b, \xi d | \lambda \zeta \xi^{-1})
		\left( \frac{(1 + \xi z)(1 + \xi z^{-1})}{1 - q}\right)^L,
		\label{eq:ZpartContour}
	\end{equation}
\end{widetext}
where for the contour of integration we take the unit circle deformed to include all poles at $z = \lambda
\zeta \xi^{-1} q^k$, $z = \xi b q^k$, and exclude all poles at $z = 1/(\lambda \zeta \xi^{-1} q^k)$, $z = 1/(\xi b
q^k)$, with $k = 0, 1, 2, \ldots$.

The deformation to include/exclude the $\lambda$ poles follows from the orthogonality condition
(Eq.~\eqref{eq:BigHorthogContour}).  The case with $b > 1$ is less straightforward. 
With $b > 1$, the product
$\langle\!\langle  h(\cos\theta) |  V \rangle\!\rangle$, which appears when we compute the partition function,
is in fact a divergent sum.  A representation without this problem is known for the single species ASEP
\cite{UchiyamaSW04}, but not in the multispecies case.  However, the deformation of the contour for the $b > 1$ case can be justified as the analytic continuation of the partition function \cite{DerridaEHP93,BlytheECE00}.

\section{Stationary properties in the thermodynamic limit}
\label{sec:phase}

For finite sizes, the integral form of the partition function $Z_L(\xi^2, \zeta)$ is difficult to work with. However, it is possible to extract its asymptotic behavior when $L$ is large, allowing the computation of stationary currents and densities.

\subsection{Phase diagram}

To find the phase diagram of the model, we need to find an asymptotic form of the partition function.
And the key to the asymptotics of the partition function are the poles due to $\lambda$, $b$ in the integral form.
For $\lambda, b < 1$, the asymptotic form can be obtained from the form Eq.~\eqref{eq:ZpartTheta} following the
method in \cite{Sasamoto99}, or by a saddle-point analysis of the complex integral (Eq.~\eqref{eq:ZpartContour})
\cite{BlytheECE00}.  For $\lambda > 1$ (or similarly $b > 1$) we must subtract the contribution of the poles
at $z = 1/(\lambda \zeta \xi^{-1} q^k)$ from this result, and add the contribution of the poles at $z =
\lambda \zeta \xi^{-1} q^k$ (see \cite{BlytheE07} for a detailed explanation).  The contribution from the
poles with $k = 0$ give the dominant asymptotic behavior.

From the asymptotic form of the partition function we can compute the species 2 current through Eq.~\eqref{eq:J2},
and the averaged densities of species 1 and 2 through Eq.~\eqref{eq:rhokdiff}.  We find three phases, as in the
model with semipermeable boundaries (see Fig.~\ref{fig:trappedphases}), which we name according to the
behavior of the species 2.

\begin{itemize}
	
	\item
	Maximum current phase (MC): For $\lambda < 1$ and $b < 1$, the asymptotic form of the partition function is
	\begin{widetext}
		\begin{equation*}
			Z_L(\xi^2, \zeta)
			\simeq
			\frac{(q; q)_\infty^3 (\zeta \lambda b, \zeta \xi^{-2} \lambda c, \zeta \lambda d; q)_\infty}
			{(\zeta \xi^{-1} \lambda, \xi b, \xi^{-1} c, \xi d; q)_\infty^2}
			\
			{_2}\phi_2 \left[
			\genfracz{\zeta \xi^{-1} \lambda, \zeta \xi^{-1} \lambda}
			{\zeta \lambda b, \zeta \lambda d} \Bigg| q, \xi^2 b d
			\right]
			\frac{[(1 + \xi)(1 + \xi^{-1})]^{3/2}}{2 \sqrt{\pi} L^{3/2}}
			\left[ \frac{(1 + \xi)^2}{1 - q} \right]^L.
	\end{equation*}
	\end{widetext}
	
From this we obtain the currents and average densities
\begin{equation*}
	J^{(2)} = \frac{1-q}{4},	\ 	\rho^{(1)} = \mathcal{O}(1/L),	\ 	\rho^{(2)} = \frac{1}{2}.
\end{equation*}
The complete leading order term of the density $\rho^{(1)}$ can be computed through Eq.~\eqref{eq:rhokdiff}, but we
have not found a simple expression for it.

\item
Low density phase (LD): For $\lambda > 1$, $\lambda > b$, the leading term comes from adding (subtracting) the
contribution of the pole at $z = \zeta \xi^{-1} \lambda$ ($z = 1/(\zeta \xi^{-1} \lambda$)), and gives

	\begin{align*}
		Z_L(\xi^2, \zeta)
		\simeq &
		\frac{(\zeta^{-2} \xi^2 \lambda^{-2}; q)_\infty}
		{(\zeta^{-1} \xi^2 b/\lambda, \zeta^{-1} c/\lambda, \zeta^{-1} \xi^2 d/\lambda; q)_\infty} \\
		& \times \left( \frac{(1 + \zeta \lambda)(1 + \zeta^{-1} \xi^2 \lambda^{-1})}{1-q} \right)^L.
	\end{align*}

From this we obtain
\begin{equation*}
J^{(2)} = \frac{(1-q) \lambda}{(1 + \lambda)^2},
\
\rho^{(1)} = \frac{\lambda - 1}{1 + \lambda},
\
\rho^{(2)} = \frac{1}{1 + \lambda}.
\end{equation*}

\item

High density phase (HD): For $b > 1$, $b > \lambda$, the leading term comes from adding (subtracting) the
contribution of the pole at $z = \xi b$ ($z = 1/(\xi b$)), and gives
	\begin{equation*}
		\begin{aligned}
		Z_L(\xi^2, \zeta)
		\simeq
		& \frac{(\zeta \xi^{-2} \lambda c, \zeta \lambda d, \xi^{-2} b^{-2}; q)_\infty}
		{(\zeta \xi^{-2} \lambda / b, b c, \xi^{-2} c/b, \xi^2 b d, d/b; q)_\infty}\\
		\times {_1}\phi_1 & \left[\genfracz{\zeta \xi^{-2} \lambda / b}{\zeta \lambda d} \Bigg| q, \xi^2 b d \right] \left( \frac{(1 + \xi^2 b)(1 + b^{-1})}{1-q} \right)^L.
		\end{aligned}
	\end{equation*}

From this we obtain
\begin{equation*}
J^{(2)} = \frac{(1-q) b}{(1 + b)^2},
\
\rho^{(1)} = \mathcal{O}(1/L),
\
\rho^{(2)} = \frac{b}{1 + b}.
\end{equation*}
Again, we have not found a simple expression for the density $\rho^{(1)}$.
\end{itemize}
\begin{figure}[]
	\centering
	\includegraphics[width=0.45\textwidth]{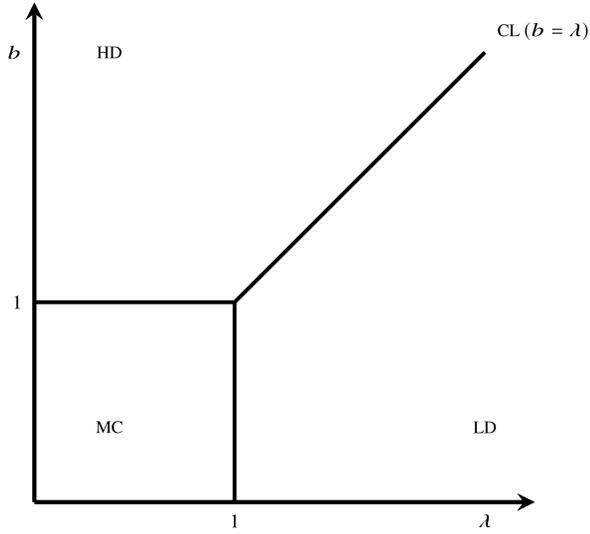}
	\caption{
		Phase diagram for lattice average current and density with $\lambda = \gamma/\alpha$.
	}
	\label{fig:phasediagram}
\end{figure}

\begin{figure*}[htbp!]
			\centering
				\includegraphics[width=0.7\textwidth]{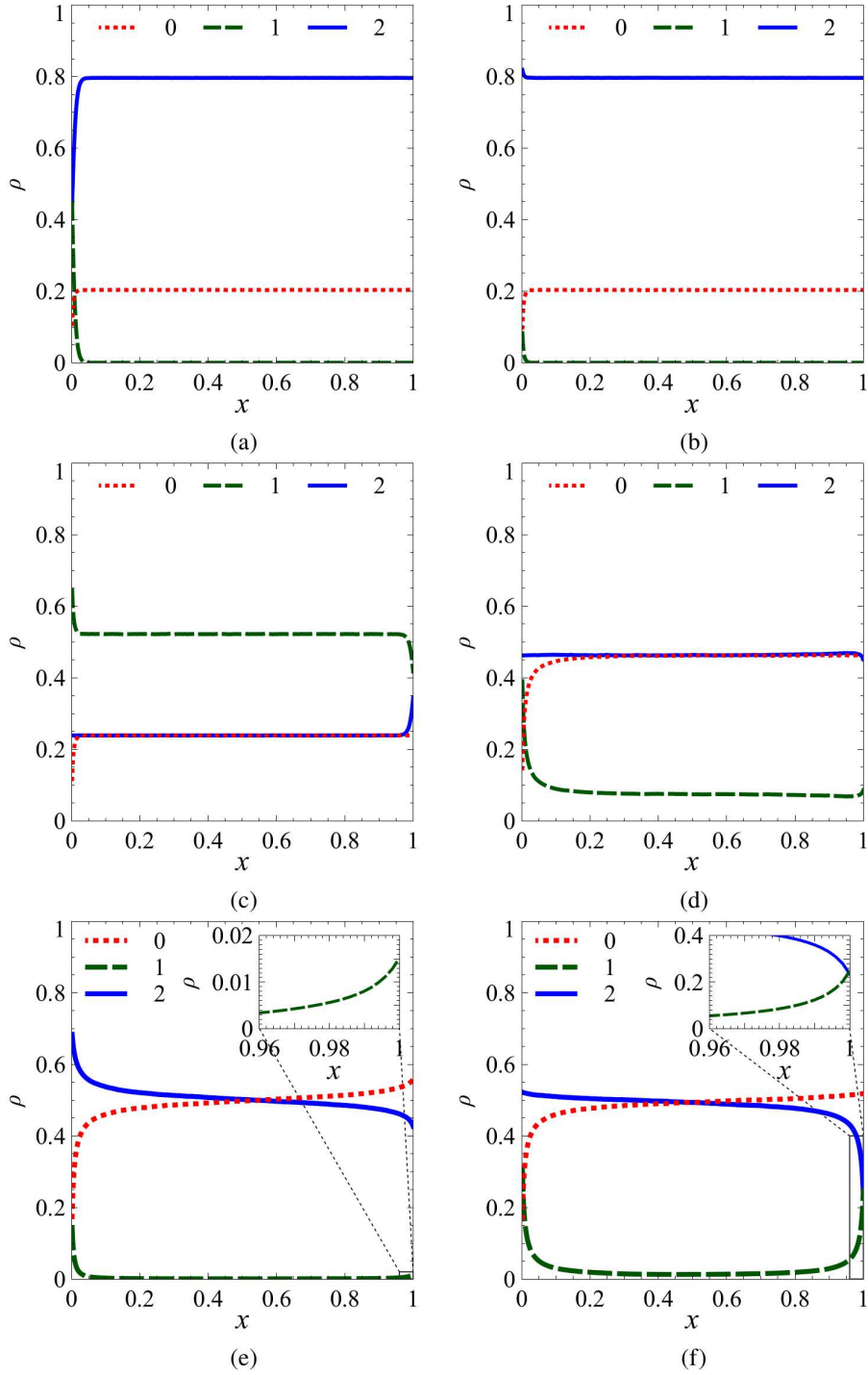}
			
			\caption{(Color online) 
				Density profiles in the LD, HD and MC phases.  Each plot shows densities of species 0 (red
				dotted line), 1 (green dashed line) and 2 (blue solid line) versus normalized site position $x=i/L$ for
				$L=500$. The sub-phases LD1, LD2, HD1 and HD2 will be described in Sec.~\ref{sec:21densities}.}
			\label{fig:simresults}
\end{figure*}

The phase diagram is shown in Fig.~\ref{fig:phasediagram}.  Simulation results showing typical density profiles for each of the phases are shown in Fig.~\ref{fig:simresults}.  The sub-phases identified in those figures will be discussed in Sec.~\ref{sec:21densities}. In each phase, the current can be
expressed in the mean-field form, $J^{(2)} = (1-q) \rho^{(2)} (1 - \rho^{(2)})$.
This is not obvious from the definition of the model because although $2$'s cannot distinguish between $0$'s and $1$'s in the bulk or at the left boundary, they can be distinguished at the right boundary.
		
\begin{figure*}[htbp!]
	\centering
	
	\includegraphics[width=0.7\textwidth]{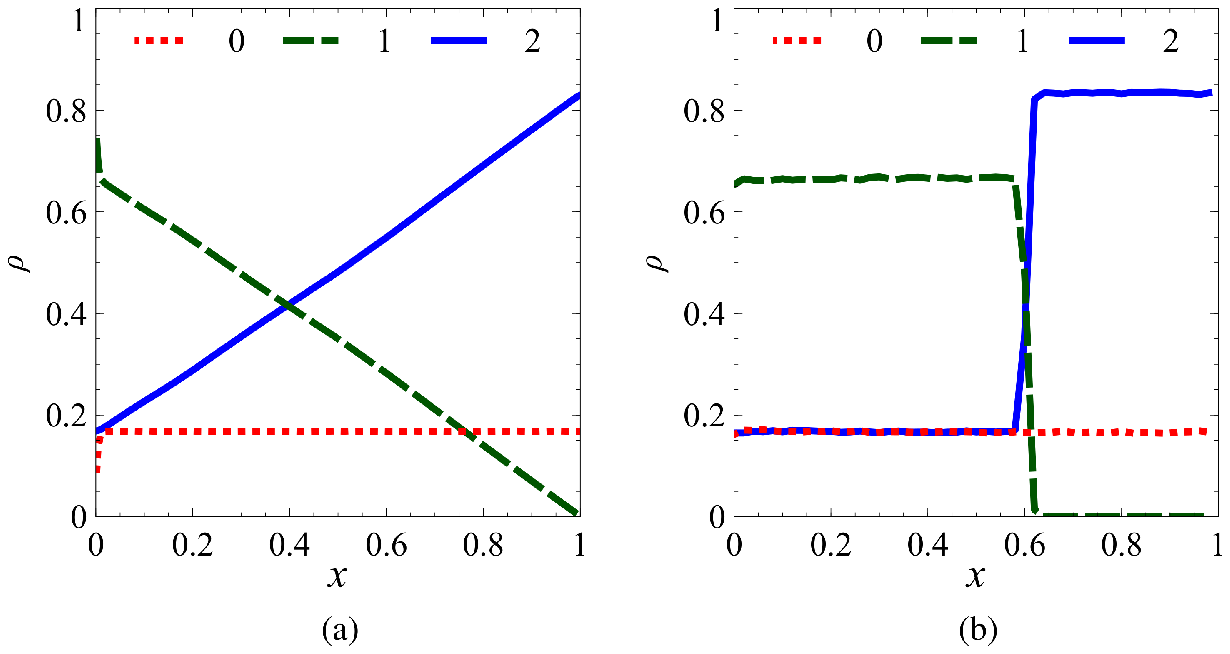}
		
	\caption{(Color online)
		Time-averaged and instantaneous density profiles on the coexistence line for $\lambda=b=4.93,d=-0.59$
		($\alpha= 0.15, \gamma =0.74 , \beta=0.28, \delta=0.89, q=0.41$).  Each plot shows densities of species 0
		(red dotted line), 1 (green dashed line) and 2 (blue solid line) versus normalized site position $x=i/L$.
	}
	\label{fig:simCLs}
\end{figure*}

As in the ASEP, there is a first order phase transition along the coexistence line (CL): that is, the line 
$\lambda = b > 1$ separating the high and low density phases.  On this line, high and low density domains
coexist on the lattice.  The left boundary attempts to impose a region characteristic of the low density phase,
while the right boundary attempts to impose a region as in the high density phase.  These two domains are
separated by a moving shock, or domain wall \cite{KolomeiskySKS98, PopkovSchutz2003}.  The linear profiles shown in
Fig.~\ref{fig:simCLs}(a) are characteristic of this situation when the position of the shock is averaged across
the lattice.  In Fig.~\ref{fig:simCLs}(b), we show an instantaneous density profile in this phase, with
the shock captured at around $0.6L$. Figure~\ref{fig:simCLs}(b) was obtained by taking a very large lattice
length ($L = 2500$), then computing a coarse-grained spatial density by averaging over windows of 50 sites.

\subsection{Boundary densities}
\label{sec:21densities}

The simulation results show that the per-site density differs from the value
averaged across the lattice.  We can get some indication of this behavior by computing the site densities
\begin{equation}
\rho^{(k)}_i = \frac{1}{Z_L} \bbra{W}C^{i-1} X_k C^{L-i}\kket{V},
\end{equation}
for species $k = 1, 2$ at the first and last sites.
We can use the boundary algebra relations (Eq.~\eqref{eq:boundaryrels}) to express the
density at site $1$ in terms of the current $J^{(2)}$ (Eq.~\eqref{eq:J2}).  We obtain
\begin{equation}
\begin{aligned}
\rho^{(1)}_1 & = \frac{-c \lambda (1-q) + \lambda(1 + c)^2 J^{(2)}}
{(1 - q)(1 - c \lambda)},
\\
\rho^{(2)}_1 & = \frac{1 - q - (1 + \lambda)(1 + c) J^{(2)}}
{(1 - q)(1 - c \lambda)}.
\end{aligned}
\end{equation}

Taking the value of $J^{(2)}$ for each phase gives the density at the left boundary (see
Table~\ref{table:densities}).
At the right boundary we find
\begin{equation}
	\begin{aligned}
		\rho^{(2)}_L = & \frac{(1 + b)(1 + d)J^{(2)} - (1 - q)b d}
		{(1 - q)(1 - b d)}  + \frac{b d}{1 - b d} \rho^{(1)}_L,
		\label{eq:rho2L}
	\end{aligned}
\end{equation}
but the algebraic relations alone are not enough to give $\rho^{(1)}_L$.  Instead we must return to the
representation of the algebra.  Again, we will use the trick of inserting the identity operator
(Eq.~\eqref{eq:hhidentity}), but now we take $\zeta = \xi = 1$, and write
\begin{widetext}
\begin{equation*}
\begin{aligned}
I_{L,k} & = \bbra{W} C^{L-k} A^k \kket{V}
 = 
\int_0^\pi \frac{\dd \theta}{2 \pi} w(\cos\theta; \lambda)
\bbra{W} C^{L-k} \kket{h(\cos\theta)} \bbra{h(\cos\theta)} A^k \kket{V},
\end{aligned}
\end{equation*}
\end{widetext}
which will allow us to compute the probability that the $k$ rightmost sites are occupied by particles of
species $1$.
Using the representation of the algebra, and the sum formula (Eq.~\eqref{eq:sumHF}), we find
\begin{widetext}
	\begin{equation*}
		\begin{aligned}
		\bbra{h(\cos\theta)} A^k \kket{V}
		& =
		\frac{\lambda^k}{(1-q)^k} \sum_{n=0}^\infty \frac{q^{kn}}{(q; q)_n} H_n(\cos\theta; \lambda | q) F_n(b, d; 0)
		 = \frac{\lambda^k}{(1-q)^k} \Theta(\cos \theta; q^k b, q^k d | \lambda).
		\end{aligned}
	\end{equation*}
\end{widetext}

Thus we find that the integral expression for $I_{L,k}$ is (up to an overall factor) simply that of the
partition function at length $L-k$ with $b \to q^k b$, $d \to q^k d$.  That is,
\begin{equation*}
I_{L,k} = \frac{\lambda^k}{(1-q)^k} Z_{L-k}\Bigg|_{\substack{b \to q^k b \\ d \to q^k d}},
\end{equation*}
where $Z_L = Z_L(\xi^2 = 1, \zeta = 1)$.

The asymptotic behavior of $I_{L,k}$ is the same as that for the partition function, except that the phase boundaries now depend on $q^k b$ instead of on $b$. We will write $Z_L^{\text{LD}}$, $Z_L^{\text{HD}}$, $Z_L^{\text{MC}}$, to indicate the expression for the partition function in the low density, high density, or maximum current phases respectively.  Then we write $I_{L,k}^{\text{XX}}$ for the corresponding value of $I_{L,k}$, but with the phase boundaries determined by $q^k b$.

\begin{figure}[ht]
	\centering
	\includegraphics[width=0.4\textwidth]{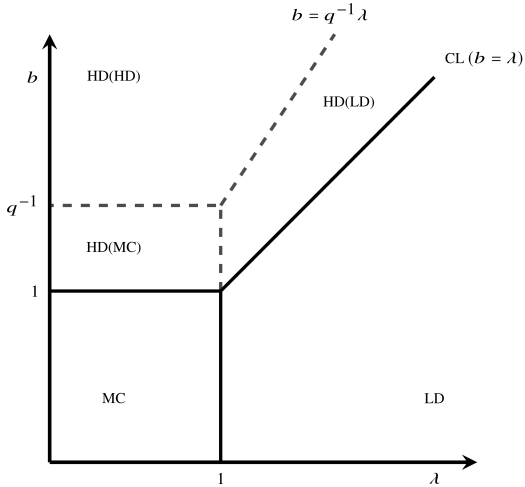}
	\caption{
		Division of the HD phase according to the scaling of $\rho^{(1)}_L$, as given in
		Eq.~\eqref{eq:rho1LHDMC} -- Eq.~\eqref{eq:rho1LHDLD}.
	}
	\label{fig:phasediagrams_scaling}
\end{figure}

Now, the probability of having the $k$ rightmost sites occupied by particles of type 1 is
\begin{equation}
P_{\text{jam}}^{(1)}(k) = P(\tau_{L-k+1} = \ldots \tau_L = 1) = \frac{I_{L,k}^{\text{YY}}}{Z_L^{\text{XX}}}.
\label{eq:P1jam}
\end{equation}
Here XX, YY indicates the appropriate phase for each part of the expression: the XX phase boundaries are
determined by $(\lambda, b)$, and the YY boundaries are determined by $(\lambda, q^k b)$.  The density
$\rho^{(1)}_L$ is given by Eq.~\eqref{eq:P1jam} with $k = 1$.  We compute $\rho^{(1)}_L$ for each possible phase
combination, indicating these by the shorthand XX(YY):
\begin{itemize}
	\item
	MC(MC) phase: $\lambda, b < 1$:
	\begin{widetext}
		\begin{equation}
			\rho^{(1)}_L = \frac{\lambda}{4} \frac{(1-b)^2(1-d)^2}{(1-\lambda b)(1-\lambda d)}
			\left(\frac{L}{L-1}\right)^{3/2}
			{_2}\phi_2\left[\genfracz{\lambda, \lambda}
			{q \lambda b, q \lambda d} \Bigg| q, q^2 b d \right]
			\Bigg/
			{_2}\phi_2\left[\genfracz{\lambda, \lambda}
			{\lambda b, \lambda d} \Bigg| q, b d \right].
			\label{eq:rho1LMC}
		\end{equation}
	\end{widetext}
	Note that $\rho^{(1)}_L$ approaches a constant value for large $L$.  Note also that for $q, |bd| \ll 1$, the
	$_2 \phi_2$ series in this expression are close to $1$, and we can approximate
	\begin{equation*}
	\rho^{(1)}_L \simeq \frac{\lambda}{4} \frac{(1-b)^2(1-d)^2}{(1-\lambda b)(1-\lambda d)}.
	\end{equation*}
	
	\item
	HD(MC) phase: $\lambda < 1$, $1 < b < q^{-1}$:
	\begin{equation}
	\rho^{(1)}_L \sim \frac{1}{(L-1)^{3/2}} \left(\frac{4}{(1 + b)(1 + b^{-1})}\right)^L.
	\label{eq:rho1LHDMC}
	\end{equation}
	By `$\sim$' we mean the scaling behavior with $L$.  We do not write out the full expression, only because we
	have not found a simple form for it.
	
	\item
	HD(HD) phase: $b > q^{-1}$, $b > q^{-1} \lambda$:
	\begin{equation}
	\rho^{(1)}_L \sim \left(\frac{(1 + q b)(1 + q^{-1} b^{-1})}{(1 + b)(1 + b^{-1})}\right)^L.
	\label{eq:rho1LHDHD}
	\end{equation}
	
	\item
	HD(LD) phase: $\lambda > 1$, $\lambda < b < q^{-1} \lambda$:
	\begin{equation}
	\rho^{(1)}_L \sim \left(\frac{(1 + \lambda)(1 + \lambda^{-1})}{(1 + b)(1 + b^{-1})}\right)^L.
	\label{eq:rho1LHDLD}
	\end{equation}
	
	\item
	LD(LD) phase: $\lambda > 1$, $\lambda > b$:
	\begin{equation}
	\rho^{(1)}_L \simeq \frac{(\lambda - b)(\lambda - d)}{(1 + \lambda)^2}.
	\end{equation}
\end{itemize}
Taking $\rho^{(1)}_L$ as the order parameter, the high density phase splits into sub-phases according to the
scaling behavior.  However across all these high density sub-phases, the density $\rho^{(1)}_L$ scales as
$z^L$ or $z^L/L^{3/2}$ with $z < 1$.  These sub-phases are depicted in Fig.~\ref{fig:phasediagrams_scaling}

The maximum current and low density phases do not split into sub-phases, and the leading order behavior is
constant in $L$.  The expressions for $P_{\text{jam}}^{(1)}(k)$ in these phases are non-vanishing (with $L$):
\begin{itemize}
	\item
	
	MC(MC) phase: $\lambda, b < 1$:
	\begin{widetext}
		\begin{equation*}
			P_{\text{jam}}^{(1)}(k) = \left(\frac{\lambda}{4}\right)^k \frac{(b, d; q)_k^2}{(\lambda b, \lambda d; q)_k}
			\left(\frac{L}{L-k}\right)^{3/2}
			{_2}\phi_2\left[\genfracz{\lambda, \lambda}
			{q^k \lambda b, q^k \lambda d} \Bigg| q, q^{2k} b d \right]
			\Bigg/
			{_2}\phi_2\left[\genfracz{\lambda, \lambda}
			{\lambda b, \lambda d} \Bigg| q, b d \right].
		\end{equation*}
	\end{widetext}
	
	\item
	LD(LD) phase: $\lambda > 1$, $\lambda > b$:
	\begin{equation*}
	P_{\text{jam}}^{(1)}(k) \simeq \frac{\lambda^k (\lambda^{-1}b, \lambda^{-1}d; q)_k}{(1 + \lambda)^{2k}}.
	\end{equation*}
\end{itemize}

\begin{table*}[htbp!]
	\centering
	\renewcommand{\arraystretch}{1.5}
	\begin{tabular}{| l | c c | c c|}
		\hline
		Phase
		& $\rho^{(1)}_1 - \rho^{(1)}$
		& $\rho^{(2)}_1 - \rho^{(2)}$
		& $\rho^{(1)}_L - \rho^{(1)}$
		& $\rho^{(2)}_L - \rho^{(2)}$
		\\
		\hline
		MC ($\lambda, b < 1$)
		& $\frac{(1-c)^2}{4(1 - c \lambda)}$
		& $\frac{(1-c)(1 - \lambda)}{4(1 - c \lambda)}$
		& $\rho^{(1)}_L$
		& $-\frac{(1-b)(1-d)}{4(1 - b d)} + \frac{b d}{1 - b d} \rho^{(1)}_L$
		\\
		LD ($\lambda > 1$, $\lambda > b$)
		& $\frac{1-c\lambda}{(1 + \lambda)^2}$
		& 0
		& $-\frac{b(\lambda - d) - (1 - d \lambda)}{(1 + \lambda)^2}$
		& $\frac{b(\lambda - d) - (1 - d \lambda)}{(1 + \lambda)^2}$
		\\
		HD ($b > 1$, $b > \lambda$)
		& $\frac{(b - c)(1- b c)\lambda}{(1 + b)^2 (1 - c \lambda)}$
		&  $\frac{(b - c)(1- b \lambda)}{(1 + b)^2 (1 - c \lambda)}$
		& 0
		& 0 \\
		\hline
	\end{tabular}
	\caption{Difference between boundary density and average bulk density in each phase.  The value $\rho^{(1)}_L$
		in the MC phase is given in Eq.~\eqref{eq:rho1LMC}.  Recall that $-1 < c,d \le 0$ in all phases.}
	\label{table:densities}
\end{table*}

\begin{figure}[ht]
	\centering
	\includegraphics[width=0.4\textwidth]{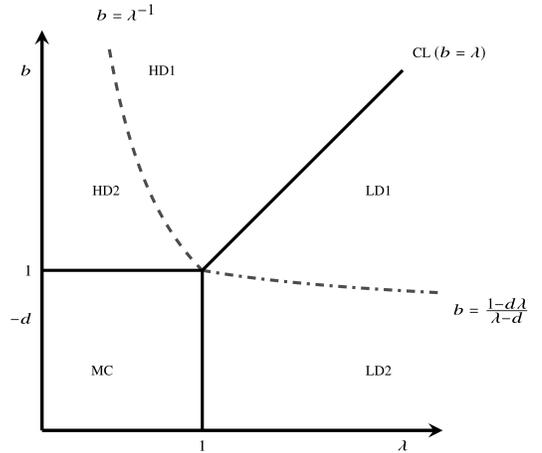}
	\caption{
		Division of phases according to the difference between bulk and boundary densities for a fixed value of $d$.
	}
	\label{fig:phasediagrams_boundaries}
\end{figure}

It might seem surprising at first glance to find that in the MC phase, species 1 has a fixed finite density at the right boundary for large $L$, even as the bulk density vanishes as $1/L$ (see Eq.~\eqref{eq:rho1LMC}). However, this can be understood from the mean field-like behavior of the system. There are only isolated $1$'s in the bulk, which perform independent asymmetric random walks with forward hopping rate $\rho^{(0)} + q \rho^{(2)}$ and reverse hopping rate 
$\rho^{(2)} + q \rho^{(0)}$. In the bulk, these are equal, but on the right boundary, $\rho^{(0)} > \rho^{(2)}$, which causes a drift towards the right leading to a buildup of $1$'s.
The insets in Fig.\ref{fig:simresults}(e), \ref{fig:simresults}(f) show close-ups of such density profiles.  The simulation results and analytically calculated values are in good agreement.

In Table~\ref{table:densities} we give the densities of species 1 and 2 at the first and last sites.  We
present these as the difference from the bulk density, i.e. $\rho^{(j)}_i - \rho^{(j)}$.
Of note is that this density difference can change signs for both species at the right boundary in the LD
phase, and for species 2 at the left boundary in the HD phase.  We identify the following subphases:
\begin{itemize}
	\item
	HD1: $b > \lambda^{-1}$, $b > \lambda$: Here $\rho^{(2)}_1 < \rho^{(2)}$.  A typical profile is shown
	in Fig.~\ref{fig:simresults}(a).
	
	\item
	HD2: $1 < b < \lambda^{-1}$: Here $\rho^{(2)}_1 > \rho^{(2)}$.  A typical profile is shown
	in Fig.~\ref{fig:simresults}(b).
	
	\item
	LD1: $\frac{1 - d \lambda}{\lambda - d} < b < \lambda$: Here $\rho^{(1)}_L < \rho^{(1)}$ and
	$\rho^{(2)}_L > \rho^{(2)}$.  A typical profile is shown in Fig.~\ref{fig:simresults}(c).
	
	\item
	LD2: $\lambda > 1$, $b < \frac{1 - d \lambda}{\lambda - d}$: Here $\rho^{(1)}_L > \rho^{(1)}$ and
	$\rho^{(2)}_L < \rho^{(2)}$.  A typical profile is shown in Fig.~\ref{fig:simresults}(d).
	
\end{itemize}

These subdivisions are depicted in Fig.~\ref{fig:phasediagrams_boundaries}.

Beyond the density difference at the boundary, it would be interesting to calculate the correlation lengths,
that is the form of the decay to the bulk density values.  For the single species ASEP, this has been studied
through a variety of different approaches
\cite{Krug91,DerridaDM92,SchuetzD93,DerridaEHP93,EsslerR96,Sasamoto99,KolomeiskySKS98}.

We also note that the phase diagrams in Fig.\ref{fig:phasediagrams_scaling} and
\ref{fig:phasediagrams_boundaries} resemble those that have appeared in other contexts in connection with the
single species ASEP.  The subdivisions of the high density phase in Fig.~\ref{fig:phasediagrams_scaling}
(related to the scaling of the density $\rho^{(1)}_L$) appear in the phase diagram for the correlation lengths
of the ASEP \cite{Sasamoto99}.  And the subdivisions in Fig.~\ref{fig:phasediagrams_boundaries} are similar (but not
identical) to the parameter constraints for which there are finite dimensional representations of the matrix
product algebra \cite{MallickS97}.  It would be interesting to know if any deeper connection exists in these
cases.

\section*{Acknowledgement}

Our warm thanks go to Matthieu Vanicat and Luigi Cantini for discussions and suggestions.
The first and third authors were partially supported by the UGC Centre for Advanced Studies. The first author
acknowledges support from the Fondation Sciences Math\'e\-matiques de Paris for a visit to IHP and from DST
grant DST/INT/SWD/VR/P-01/2014.  The second author acknowledges partial support from the grant AAP MASHE for
USMB (Universit{\'e} Savoie Mont Blanc).
We also thank the anonymous referees for useful comments.

\end{document}